\documentstyle[12pt]{article}
\begin{document}
{\large
%%%%%%%%%%%%%%%%%%%%%%%%%%%%%%%%%%%%%%%%%%%%%%%%%%%%%%%%%%%%%%%%%%%%%%%%%%
\begin{center}
{\Large \bf
{FERMION SCATTERING BY PHASE BOUNDARIES }
}\\
\vskip 1.5cm
{ Andro BARNAVELI\footnote{e-mail: barbi@iberiapac.ge}
and Merab GOGBERASHVILI\footnote{e-mail: gogber@physics.iberiapac.ge} } \\
\vskip 0.3cm
{\it
 {Institute of Physics of the Georgian Academy of Sciences,}\\
 {Tamarashvili str. 6, Tbilisi 380077, Republic of Georgia.}
} \\
\vskip 2.0cm
%%%%%%%%%%%%%%%%%%%%%%%%%%%%%%%%%%%%%%%%%%%%%%%%%%%%%%%%%%%%%%%%%%%%%%%%%%%
{\bf Abstract}\\
\vskip 0.5cm
\quotation
{\small
We consider interactions of fermions 
with the domain wall bubbles produced during the first order phase 
transitions. New exact solution of Dirac equations and reflection 
coefficient are obtained.} \end{center}
\vskip 2.0cm
%%%%%%%%%%%%%%%%%%%%%%%%%%%%%%%%%%%%%%%%%%%%%%%%%%%%%%%%%%%%%%%%%%%%%%%%%%%%

\rm
During these last few years, a considerable amount of works has been 
devoted to the analysis of possibility of generation of baryon asymmetry 
of the universe on the electro-weak scale (see for example \cite{ckn-93, 
fs-93, ckn-92}). In these models usually considered first order phase 
transitions, which can be described in terms of
nucleation of bubbles of "true" vacuum (with an inner vacuum expectation
value of Higgs field $v \neq 0$) appearing in the preexisting "false"
vacuum (with $v = 0 $ throughout) and growing until the Universe is
completely converted to the broken phase. So it is important to 
investigate interaction of particles from cosmological plasma with new phase 
bubbles.  In this paper 
the new exact solution of Dirac equations is obtained for realistic bubble 
wall potential.

The tree potential of the Higgs field has the form:
\begin{equation}
 V = \frac{\lambda}{4}\varphi ^2(\varphi^2 - v^2),
\label{1}
\end{equation}
where $\lambda$ is the self-coupling constant of the scalar field.  
Once the new phase bubble has grown to macroscopic size 
the curvature of its wall can be neglected and can be considered as a 
flat. The world is divided in two zones, on the left-hand side of the 
wall, say, $v = 0$, while
on the right-hand side --- $v \neq 0$, and, thus due to Yukawa coupling 
quarks and leptons acquire their masses. In this frame it is assumed that the
bubble expands from right to left. The
analytic solution of Higgs field equation for a bubble wall with its normal
along the $z$-axis and position at $z=o$ is \cite{ckn-93}:
\begin{equation}
\varphi (z) =
\frac{v}{2}\left[ 1 + th\left( \frac{z}{\delta } \right) \right] ,
\label{2}
\end{equation}
where
\begin{equation}
\delta = \frac{\sqrt{8}}{v\sqrt{\lambda }} = \frac{\sqrt{2}}{M_H}~~ ,
\label{3}
\end{equation}
is the width of the wall and $M_H = \sqrt{\lambda }v/2$ is the
Higgs mass.  We want to emphasize that the domain wall profile (\ref{2}) 
follows
directly from the field equations for the Higgs field in difference to
various "written by hand" non realistic wall-profiles which have been
considered \cite{fs-93,ckn-92}.

For the simplicity let us consider only one fermion flavor. The fermion mass
\begin{equation}
M(z) = g\varphi (z) =
\frac{m}{2}\left[ 1 + th\left( \frac{z}{\delta } \right) \right] 
\label{4}
\end{equation}
(where $g$ is Yukawa coupling constant) is a function of position with the
boundary conditions
\begin{equation}
M(-\infty ) = 0 ; \quad M(\infty ) = gv = m ,
\label{5}
\end{equation}
i.e. it is zero outside and nonzero inside the bubble.

Let us now consider scattering problem.
Assume that incident particle, coming from the left, in the domain wall frame
interacts with the wall. This puzzle is very much like the quantum mechanical
scattering of Dirac particles from a potential barrier \cite{f}. The
difference from the quantum mechanical case is that an exterior field
appears in the mass of scattered particle. Thus we are solving Dirac
equation with varying mass. In order to extract the reflection coefficients
we need to find plane wave solutions corresponding to incoming and outgoing
particles moving along $z$-axis. The components of the momentum parallel to
the wall are conserved in the scattering process and are unimportant in a
qualitative discussions. In order to solve the problem of reflection from
the wall one has to specify the boundary conditions. Particles moving in the
unbroken phase have a wave function proportional to
\begin{equation}
\psi \sim e^{-iEt \pm iqz},
\label{6}
\end{equation}
where the upper sign corresponds to incident particle moving towards the
wall, while the lower sign --- to the reflected particle. The wave function
of particles transmitted through the wall is proportional to
\begin{equation}
\psi \sim e^{-iEt + ipz} .
\label{7}
\end{equation}
Here
$p=\sqrt{E^2-m^2}$ , $q= |E|$.
The boundary conditions for antiparticles are the similar as for particles.

We shall work in chiral basis (the notation of $\gamma$-matrixes the same as
in Ref. \cite{ckn-92} ). For the particles with positive energy Dirac equation
with a position-dependent mass (\ref{4}) can be written in components in the 
form \begin{eqnarray}
i\frac{d}{dz}\psi_1 + E\psi_1 + M(z)\psi_3 = 0 , \nonumber \\
i\frac{d}{dz}\psi_3 - E\psi_3 - M(z)\psi_1 = 0 ,
\label{8}
\end{eqnarray}
where $\psi_1$ and $\psi_3$ correspond to the left- and the right- handed
particles.

Introducing a new variable
\begin{equation}
y \equiv \frac{1}{1+e^{2z/\delta}}
\label{9}
\end{equation}
and searching the solution of (\ref{8}) in the form
\begin{equation}
\psi_{1(3)} \equiv y^{-i\delta p/2}(1-y)^{\pm i\delta E/2}\eta _{1(3)}(y) ,
\label{10}
\end{equation}
(where the upper sign in the second exponent corresponds to $\psi_1$, while
the lower one --- to $\psi_3$) one yields the equation:
\begin{eqnarray}
\{ y(1-y)\frac{d^2}{dy^2}+[(1-i\delta p)-
(1-i\delta p \pm i\delta E)y]\frac{d}{dy}- \\ \nonumber      
- \frac{1}{2}\delta^2E(E \mp p) \} \eta_{1(3)} = 0 . ~~~~~~~~~~~~~~~~~~~
\label{11}
\end{eqnarray}
(We want to note that in the first exponent in (\ref{10}) we could choose the
opposite sign, however in that case we would yield the outcoming plane wave
at $+\infty$ with incorrect sign of momentum $p$). Equation (\ref{11}) is
the one for Hypergeometric functions \cite{gr} and has a solutions \cite{f}:
\begin{equation}
\eta_{1(3)} = F\left[ \frac{\delta}{2}( m - ip \pm iE ) ,
\frac{\delta}{2}(-m -ip \pm ie ) , 1 - i\delta p , y \right] .
\label{12}
\end{equation}

Now let us consider an asymptotical behavior of functions $\psi_1$ and
$\psi_3$. In the area $z \rightarrow -\infty$, i.e. when $y \rightarrow 1$
and $(1-y) \rightarrow e^{2z/\delta}$, using the features of hypergeometric
functions we obtain:
\begin{equation}
\psi_{1(3)} \rightarrow A_{1(3)}e^{\pm iEz} .
\label{13}
\end{equation}
Thus $\psi_1$ corresponds to the incident (coming from the left) left handed
massless particle with momentum equal to $|E|$ while $\psi_3$ --- to the
reflected particle of the right chirality.

Constants $A_{1(3)}$  are expressed by Gamma functions in the
following way:
\begin{equation}
A_{1(3)} = \frac{\Gamma (1-i\delta p)\cdot
\Gamma (1\pm i\delta E)}{\Gamma \left[ \frac{\delta}{2}(-m-ip\mp
iE)+1\right]\cdot\Gamma \left[ \frac{\delta}{2}(m-ip\mp iE)+1\right] }~~ .
\label{14}
\end{equation}

In the broken phase, when $z \rightarrow \infty $, i.e. when $y \rightarrow
e^{-2z/\delta}, (1-y) \rightarrow 1$, we have only the wave corresponding to
transmitted massive particle:
\begin{equation}
\psi_{1(3)} \rightarrow e^{ipz}
\label{15}
\end{equation}
with momentum $p=\sqrt{E^2-m^2}$.

To find the reflection coefficient we must compute the ratio of reflected
flux of particles to the incident one. Substituting (\ref {13}) into 
expression for flux we yield:
\begin{equation}
J^z = i\psi^\dagger\gamma^0\gamma^3\psi =
\psi_1^\ast\psi_1 - \psi_3^\ast\psi_3 = |A_1|^2 - |A_3|^2 =
J_{ins}^z - J_{ref}^z .
\label{16}
\end{equation}
The reflection coefficient is:
\begin{equation}
R = \frac{J^z_{ref}}{J^z_{ins}} = \left|{A_3} \over {A_1} \right|^2 .
\label{17}
\end{equation}
Substituting expression (\ref{14}) in this formula and using the features of 
Gamma functions \cite{gr}:
\begin{eqnarray}
 \Gamma (1+X+iY)\cdot\Gamma (1+X-iY)\cdot\Gamma (1-X+iY)\cdot
 \Gamma (1-X-iY) = \\ \nonumber
= \frac{2\pi^2(X^2+Y^2)}{sh(2\pi Y) - cos(2\pi X)}~~ ,~~~~~~~~~~~~~~~~~~~~~~~
\label{18}
\end{eqnarray}
we have
\begin{equation}
R = \frac{E+p}{E-p}\cdot\frac{sh[\delta\pi (E-p)] - cos(\delta\pi m)}{sh[
\delta\pi (E+p)] - cos(\delta\pi m)}~~ .
\label{19}
\end{equation}

Thus in this letter we derived an analitical expression of the 
reflection coefficient. This expression can be 
used, for example, in  calculations of baryon asymmetry of the universe.
\vskip 1cm
The research described in this publication was made possible in part by
Grant MXL000 from the International Science Foundation.

%%%%%%%%%%%%%%%%%%%%%%%%%%%%%%%%%%%%%%%%%%%%%%%%%%%%%%%%%%%%%%%%%%%%%%%

\end{document}